\documentclass{jnmp}

\def \Oeuvres{O$\!$euvres}
 
\def\LHS{l.h.s.~}

\def \ccomma{\raise 2pt\hbox{,}} 
\def \Log {\mathop{\rm Log}\nolimits}
\def\CRAS{C.~R.~Acad.~Sc.~Paris}
\def\Pn     {{\rm Pn}}
\def\PII    {{\rm P2}}
\def\PIII   {{\rm P3}}
\def\PIV    {{\rm P4}}
\def\PV     {{\rm P5}}
\def\PVI    {{\rm P6}}

\def\Alpha{A}

\def\Beta {B}
\def\abcd{\alpha,\beta,\gamma,\delta}
\def\ABCD{\Alpha,\Beta,\Gamma,\Delta}

\def\veca{\mbox{\boldmath{$\alpha$}}}
\def\vecA{\mbox{\boldmath{$\Alpha$}}}
\def\vect{\mbox{\boldmath{$\theta$}}}
\def\vecT{\mbox{\boldmath{$\Theta$}}}

\def\BiT{birational transformation}

\def \TPVb    {{\rm T}_{\rm 5,b}}   
\def \TPVu    {{\rm T}_{\rm 5,u}}   

\def\Sa  {{\rm S}_{\rm a}}
\def\Sb  {{\rm S}_{\rm b}}
\def\Sc  {{\rm S}_{\rm c}}

\def \H {{\rm H}}  

\JNMPnumberwithin{equation}{section}

\theoremstyle{definition}

\begin{document}

%
\renewcommand{\evenhead}{Robert Conte and Micheline Musette}
\renewcommand{\oddhead}
 {A truncation for all the first degree birational transformations \dots}

%
\thispagestyle{empty}

\FirstPageHead{200*}{*}{*}{\pageref{firstpage}--\pageref{lastpage}}{Article}

\copyrightnote{200*}{Robert Conte and Micheline Musette}

\Name
 {A truncation for obtaining all the first degree birational transformations 
  of the Painlev\'e transcendents}

\label{firstpage}

\Author{Robert CONTE~$^\dag$ and Micheline MUSETTE~$^\ddag$}

\Address{$^\dag$ Service de physique de l'\'etat condens\'e, CEA--Saclay
\\ F--91191 Gif-sur-Yvette Cedex, France \\
~~E-mail: Conte@drecam.saclay.cea.fr\\[10pt]
$^\ddag$ Dienst Theoretische Natuurkunde, Vrije Universiteit Brussel
\\ Pleinlaan 2, B--1050 Brussels, Belgium \\
~~E-mail: MMusette@vub.ac.be. nlin.SI/0110031}

\Date{Received 20 July 2001; Revised 28 August 2001; 
 Accepted 29 August 2001}

\begin{abstract}
\noindent
A birational transformation is one which leaves invariant an ordinary
differential equation, only changing its parameters.
We first recall the consistent truncation
which has allowed us to obtain the first degree {\BiT} of Okamoto
for the master Painlev\'e equation $\PVI$.
Then we improve it by adding a preliminary step,
which is to find all the Riccati subequations of the considered $\Pn$ 
before performing the truncation.
We discuss in some detail the main novelties of our method,
taking as an example the simplest Painlev\'e equation for that purpose,
$\PII$.
Finally,
we apply the method to $\PV$ and obtain its two inequivalent first degree
{\BiT}s.
\end{abstract}

\section{Introduction} 

A \textit{birational transformation} is by definition a set of two relations,
\begin{eqnarray} 
& &
u = f(U',U,X),\
U = F(u',u,x),
\label{eqbira}
\end{eqnarray}
with $f$ and $F$ rational functions,
which maps an algebraic ordinary differential equation (ODE), 
for instance a Painlev\'e equation,
\begin{eqnarray} 
& &
E(u) \equiv \Pn(u,x,\veca)=0,\
\veca=(\abcd),\
\label{eqPnu}
\end{eqnarray}
into the same equation with different parameters
\begin{eqnarray} 
& &
E(U) \equiv \Pn(U,X,\vecA)=0,\
\vecA=(\ABCD),
\label{eqPnU}
\end{eqnarray}
with some homography (usually the identity) between $x$ and $X$.
The parameters $(\veca,\vecA)$ must obey as many algebraic relations 
as elements in $\veca$.
The \textit{degree} of a \BiT\ is defined as the highest degree in $U'$
or $u'$ 
(or more generally in the $(N-1)$th derivative of $U$ and $u$)
of the numerator and the denominator of (\ref{eqbira}).

A method allowing one to derive such {\BiT}s was recently introduced
\cite{CJP,GJP1999a},
and later improved
\cite{CM2001alreadysubmitted,CM2001c}
so as to provide {\BiT}s which have a degree equal to one
for any first degree $N$-th order ODE.
Its application to the master Painlev\'e equation $\PVI$,
\begin{eqnarray*}
\label{eqP6}
\PVI\ : \
u''
&=&
\frac{1}{2} \left[\frac{1}{u} + \frac{1}{u-1} + \frac{1}{u-x} \right] {u'}^2
- \left[\frac{1}{x} + \frac{1}{x-1} + \frac{1}{u-x} \right] u'
\\
& &
+ \frac{u (u-1) (u-x)}{x^2 (x-1)^2} 
  \left[\alpha + \beta \frac{x}{u^2} + \gamma \frac{x-1}{(u-1)^2} 
        + \delta \frac{x (x-1)}{(u-x)^2} \right]\ccomma
\end{eqnarray*}
provided the \BiT\ \cite{C2001TSP6}, 
already found by Okamoto \cite[p.~356]{Okamoto1987I},
\begin{eqnarray}
\frac{N}{u-U}
& = &
  \frac{x (x-1) U'}{U (U-1)(U-x)}
 +\frac{\Theta_0}{U}+\frac{\Theta_1}{U-1}+\frac{\Theta_x-1}{U-x}
\label{eqTP6uvecTUnsigned}
\\
& = &
  \frac{x(x-1)u'}{u(u-1)(u-x)}
 +\frac{\theta_0}{u}+\frac{\theta_1}{u-1}+\frac{\theta_x-1}{u-x}\ccomma
\label{eqTP6uvectUnsigned}
\\
\theta_j
& = &
\Theta_j - \frac{1}{2} \left(\sum \Theta_k\right) + \frac{1}{2}
\ccomma\
j,k=\infty,0,1,x,
\label{eqT6}
\\
\Theta_j
& = &
\theta_j - \frac{1}{2} \left(\sum \theta_k\right) + \frac{1}{2}
\cdot
\label{eqT9}
\end{eqnarray}
The transformation is clearly birational since the \LHS\ is homographic
in both $u$ and $U$.
In the above, the monodromy exponents
$\vect=(\theta_\infty,\theta_0,\theta_1,\theta_x)$
are defined as
\begin{eqnarray} 
& &
\theta_\infty^2= 2 \alpha,\
\theta_0^2     =-2 \beta,\
\theta_1^2     = 2 \gamma,\
\theta_x^2     =1 - 2 \delta,
\end{eqnarray}
and similarly for their uppercase counterparts,
while the odd-parity constant $N$ takes the equivalent expressions
\begin{eqnarray}
N
& = &
\sum (\theta_k^2 - \Theta_k^2)
\label{eqP6N}
\\
& = &
  1 - \sum     \Theta_k
=
 -1 + \sum     \theta_k
\label{eqNvect}
\\
& = &
  2 (    \theta_j -     \Theta_j),\ j=\infty,0,1,x.
\label{eqNvecT}
\end{eqnarray}
The choice of the eight signs of $\vect$ and $\vecT$ is such that 
the square of this transformation is the identity.
We will adopt such a convention (choice of signs so as to have involutions)
throughout the present paper.
This will dispense us from writing the second half, e.g.
(\ref{eqTP6uvectUnsigned}),
of a \BiT.
Indeed, if the \LHS\ of the first half is chosen invariant under
the permutation of $(u,\vect)$ and $(U,\vecT)$,
which is the case in (\ref{eqTP6uvectUnsigned}),
the second half is deduced from the first half by just
permuting the lowercase and uppercase notation.

The method is an extension to ODEs of the powerful
\textit{singular manifold method} introduced by Weiss \textit{et al.}
\cite{WTC}.
This method in turn mainly assumes the existence of a \textit{truncation},
i.e. a representation of the solution $u$ of the considered ODE 
(\ref{eqPnu})
by a Laurent series which terminates (``truncates'').
Its current achievements are detailed in summer school proceedings,
see Refs.~\cite{Cargese1996Musette,CetraroConte}.

The purpose of this article is threefold.
Firstly, we present a significant improvement to that method,
only based on the consideration of Riccati equations.
This improvement reduces the obtaining of the above \BiT\ of $\PVI$
to computations which are easily feasible by hand.
Secondly,
using the simplest equation $\PII$ as an example,
we point out the main differences between our method 
and the one previously introduced \cite{CJP,GJP1999a}.
Thirdly,
on the specific example of $\PV$, 
we show that the straightforward application of our method yields
the two first degree {\BiT}s of this ODE.

The organization of the paper is as follows.
In Section \ref{sectionHomographyHK},
we recall the main difference between the truncation of an ODE
and that of a partial differential equation (PDE),
which is a fundamental homography between the ODE and 
the truncation variable,
without any counterpart for a PDE.

In Section \ref{sectionOneFamilyTruncationHK},
we define the consistent truncation which implements this homography
and we present the improvement.

The next Section \ref{sectionAnElementaryExampleP2} is devoted to
a parallel processing of the simplest Painlev\'e equation $\PII$,
so as to clearly point out the differences between the previous method
and ours.

Finally, in Section \ref{sectionImprovedTruncationP5},
we process $\PV$ with our method,
and find its two inequivalent first degree {\BiT}s.

Throughout this article, we discard the nongeneric cases in which the 
components of $\veca$ are constrained.
One such case is the well known birational transformation between $\PV$
with $\delta=0$ and $\PIII$.

\section{The fundamental homography, a difference with PDEs} 
\label{sectionHomographyHK}

Consider a Painlev\'e ODE (\ref{eqPnU}) which admits a \BiT, 
i.e. $n=2,3,4,5,6$.
There exist two Riccati equations associated to this $\Pn$.
The first one is the Painlev\'e equation (\ref{eqPnU}) itself.
Indeed, any $N$-th order, first degree ODE with the Painlev\'e property
is necessarily \cite[pp.~396--409]{PaiLecons}
a Riccati equation for $U^{(N-1)}$,
with coefficients depending on $x$ and the lower derivatives of $U$,
in our case
\begin{eqnarray}
& &
U''=A_2(U,x) {U'}^2 + A_1(U,x) U' + A_0(U,x).
\label{eqRiccatiUprime}
\end{eqnarray}

The second Riccati equation is the algebraic transform for $Z=\psi/\psi'$
of the linear second order ODE for $\psi$ which has been built
by Richard Fuchs \cite{FuchsP6} for $\PVI$,
and by confluence to any other $\Pn$,
\begin{eqnarray} 
& &
Z'=1 + z_1 Z + z_2 Z^2.\ 
\label{eqRiccatiOneFamily}
\end{eqnarray}

Since the group of invariance of a Riccati equation is the homographic group,
the variables $U'$ and $Z$ are linked by a homography.
Let us define it as
\begin{eqnarray}
& &
 (U' + g_2) (Z^{-1} - g_1) - g_0=0,\ g_0 \not=0,
\label{eqHomographyRUprime}
\end{eqnarray}
or, in the affine case, as
\begin{eqnarray}
& &
(U' + G_2) - G_0 Z^{-1} =0,\ G_0 \not=0.
\label{eqLinearZUprime}
\end{eqnarray}
The coefficients $g_j$ or $G_j$ are rational in $(U,x)$.
We will not consider (\ref{eqLinearZUprime}) in the present paper.

This homography allows us to compute the two coefficients
$z_j$ of the Riccati pseudopotential equation (\ref{eqRiccatiOneFamily})
as explicit expressions 
of $(g_j, \partial_U g_j, \partial_x g_j, A_2,A_1,A_0,U')$.
Indeed,
eliminating $U'$ between (\ref{eqRiccatiUprime}) and
(\ref{eqHomographyRUprime}) defines a first order ODE for $Z$,
whose identification with (\ref{eqRiccatiOneFamily})
\textit{modulo} (\ref{eqHomographyRUprime})
provides three relations,
\begin{eqnarray} 
& &
g_0= g_2^2 A_2 - g_2 A_1 + A_0 + \partial_x g_2 - g_2 \partial_U g_2,
\label{eqg0Apriori}
\\
& &
z_1=
   A_1 - 2 g_1 + \partial_U g_2 - \partial_x \Log g_0 
+ \left(2 A_2 - \partial_U \Log g_0\right) U',
\label{eqz1Apriori}
\\
& &
z_2= -g_1 z_1 - g_1^2 - g_0 A_2 - \partial_x g_1 - (\partial_U g_1) U'.
\label{eqz2Apriori}
\end{eqnarray}
Since $A_2,A_1,A_0$ are given,
there only remains to determine the two
coefficients $g_1,g_2$ of the homography,
which are functions of the two variables $U,x$.
In particular,
the two functions $z_1,z_2$ of the three variables $U',U,x$
are not the elementary quantities to determine.

For a PDE, 
the homography (\ref{eqHomographyRUprime}) does not exist any more
since at least (\ref{eqRiccatiUprime}) does not survive.
For searching the B\"acklund transformation
(see summer schools lecture notes \cite{Cargese1996Musette,CetraroConte}
and references therein),
this makes the situation much easier since one does not have to
assume a dependence between $Z$ and some derivative of $U$.
On the contrary, for ODEs, 
if one handles $Z$ and $U'$ as independent variables,
this creates many difficulties.

\section{The improvement to the truncation} 
\label{sectionOneFamilyTruncationHK}

Each $\Pn$ equation which admits a \BiT\ has one or several
(four for $\PVI$)
couples of families of movable simple poles with opposite residues $\pm u_0$,
and the assumption for the one-family truncation is
\begin{eqnarray} 
& &
u=u_0 Z^{-1} +U,\ u_0 \not=0,\
x=X,
\label{eqDTOne}
\end{eqnarray}
in which $u$ and $U$ satisfy (\ref{eqPnu}) and (\ref{eqPnU}),
and $Z$ satisfies (\ref{eqRiccatiOneFamily}).
After determination of the rational functions $g_1(U,x)$ and $g_2(U,x)$,
the first half of the \BiT\ is
\begin{eqnarray}
& &
u=U + u_0 \left(g_1(U,x) + \frac{g_0(U,x)}{U'+g_2(U,x)}\right),
\label{eqSTDirecte}
\end{eqnarray}
with the restriction that its denominator should not vanish.
We recently conjectured \cite{CM2001alreadysubmitted}
that the ODE defined by this denominator,
\begin{eqnarray}
& &
U'+g_2(U,x)=0,
\label{eqRiccatiDenom}
\end{eqnarray}
has the Painlev\'e property,
which restricts $g_2$ to an arbitrary second degree polynomial of $U$ with
coefficients depending on $x$.

Let us prove this conjecture and completely determine $g_2$  
for $\PVI$,
and therefore for any $\Pn$ equation thanks to the confluence.

\begin{proof}
The equation (\ref{eqSTDirecte}) is equivalently written
\begin{eqnarray}
& &
\frac{u_0 g_0}{u-U-u_0 g_1} = U'+g_2,
\label{eqSTDirecteBest}
\end{eqnarray}
and the nonvanishing condition $u-U-u_0 g_1 \not=0$
does not restrict $U$ any more since the restriction concerns $u$.
Therefore the equation (\ref{eqSTDirecteBest}) still holds
when, simultaneously,
$g_0(U,x)$ vanishes and $U$ satisfies the equation (\ref{eqRiccatiDenom}).
In the case of $\PVI$,
the corresponding values $\tilde g_2$ and $\tilde g_0$ are defined
by the Riccati subequation,
\begin{eqnarray}
& &
\tilde g_2=\frac{U (U-1)(U-x)}{x (x-1)}
\left(\frac{\Theta_0}{U}+\frac{\Theta_1}{U-1}+\frac{\Theta_x-1}{U-x}\right)
\ccomma
\label{eqP6g2subeq}
\end{eqnarray}
and by the formula (\ref{eqg0Apriori}) applied to $\tilde g_2$,
\begin{eqnarray}
{\hskip -1.0truecm}
& &
\tilde g_0=(1-\Theta_\infty-\Theta_0-\Theta_1-\Theta_x)
           (1+\Theta_\infty-\Theta_0-\Theta_1-\Theta_x)
    \frac{U(U-1)(U-x)}{8 x^2(x-1)^2}\ccomma
\end{eqnarray}
which indeed defines the constraint on $\vecT$.
The couple $(g_0,g_2)$ in 
(\ref{eqSTDirecteBest}) cannot be different 
(\textit{modulo} the homographies on $U$ which preserve $x$ and $\PVI$)
from this couple $(\tilde g_0,\tilde g_2)$,
since this Riccati subequation is unique.
\end{proof}

Denoting $R$ (like Riccati) the quantity
\begin{eqnarray}
& &
R=\frac{x (x-1)U'}{U (U-1)(U-x)}
 +\frac{\Theta_0}{U}+\frac{\Theta_1}{U-1}+\frac{\Theta_x}{U-x}\ccomma
\label{eqP6NormalR}
\end{eqnarray}
one has the identity, whatever be $\vecT$,
\begin{eqnarray}
& &
x (x-1) R'
+\frac{1}{2}(1-\Theta_\infty-\Theta_0-\Theta_1-\Theta_x)
            (1+\Theta_\infty-\Theta_0-\Theta_1-\Theta_x)
\nonumber \\ & & \phantom{x R'}
-\left(
  (\Theta_1+\Theta_x-1)U+(\Theta_0+\Theta_x-1)(U-1)+(\Theta_0+\Theta_1)(U-x)
 \right) R
\nonumber \\ & & \phantom{x R'}
+\frac{1}{2}\left((U-1)(U-x)+U(U-1)+U(U-x)\right) R^2
\nonumber \\ & & \phantom{x R'}
+\frac{x^2(x-1)^2}{U(U-1)(U-x)}\PVI(U)\equiv 0,
\label{eqP6NormalIdentity}
\end{eqnarray}

To summarize, the information is twofold.
\begin{enumerate}
\item
The coefficient $g_2$ is determined 
by the fact that 
equation (\ref{eqRiccatiDenom}) must be a subequation of (\ref{eqPnU}).

\item
The coefficient $g_0$ factorizes as $g_0(U,x,\ABCD)=f_0(\ABCD) h_0(U,x)$,
defining the condition $f_0(\ABCD)=0$ for the existence of the subequation.
\end{enumerate}

The improved method is now the following.
Before performing the truncation,
one computes all the identities like (\ref{eqP6NormalIdentity})
involving Riccati subequations.
Each identity defines an explicit value for $g_2$.
For each such $g_2$, the coefficient $g_0$ is explicitly given as a factorized
expression by the formula (\ref{eqg0Apriori}).
Finally, one performs the truncation to find $g_1$ and the
algebraic relations between $\abcd$ and $\ABCD$.
Let us now describe this truncation.

The field $u$ is represented, see (\ref{eqDTOne}),
by a Laurent series in $Z$ which terminates (``truncated series'').
The \LHS\ $E(u)$ of the equation 
can similarly be written as a truncated series in $Z$.
This is achieved 
by the elimination of $u,Z',U'',U'$
between (\ref{eqPnu}), (\ref{eqPnU}), (\ref{eqDTOne}), 
(\ref{eqRiccatiOneFamily})
and (\ref{eqHomographyRUprime}),
followed by the elimination of $(g_0,z_1,z_2)$
from (\ref{eqg0Apriori})--(\ref{eqz2Apriori})
($q$ denotes the singularity order of $\Pn$ written as a 
differential polynomial in $u$, it is $-6$ for $\PVI$),
\begin{eqnarray}
& &
E(u) = \sum_{j=0}^{-q+2} E_j(U,x,u_0,g_1,\veca,\vecA) Z^{j+q-2}=0,
\label{eqST1LaurentE}
\\
& &
\forall j\ :\            E_j(U,x,u_0,g_1,\veca,\vecA)=0.
\label{eqST1Determining}
\end{eqnarray}
The nonlinear \textit{determining equations} $E_j=0$ are independent of $U'$,
and this is the main difference with previous work \cite{GJP1999a}.
Another difference is the greater number ($-q+3$ instead of $-q+1$)
of equations $E_j=0$,
which is due to the additional elimination of $U'$ with 
(\ref{eqHomographyRUprime}).

The $-q+3$ determining equations (\ref{eqST1Determining}) 
in the unknown function $g_1(U,x)$
and the unknown algebraic relations between $\abcd$ and $\ABCD$
must be solved, as usual, by increasing values of their index $j$.

\section{An elementary example: the second Painlev\'e equation} 
\label{sectionAnElementaryExampleP2} 

\begin{eqnarray*}
\PII\ : \
u''
&=&
\delta (2 u^3 + x u)
 + \alpha.
\end{eqnarray*}

The data for the two opposite families of movable singularities of $\PII$ are
\begin{eqnarray}
& &
p=-1,\
q=-3,\
u_0=d^{-1},\
\delta=d^2,\
\hbox{ Fuchs index } 4,
\end{eqnarray}
in which $d$ is any square root of $d^2$.
The unique monodromy exponent $\theta_\infty$ is defined as
\begin{eqnarray}
& &
\alpha=-d \theta_\infty.
\end{eqnarray}

Let us compare the truncation of the previous section
with the truncation introduced by \cite{CJP} and applied by \cite{GJP1999a}.
For brevity, the former will be qualified ``full'' and the latter ``semi''
for reasons to become clear soon.

\subsection{Processing of $\PII$ with the full truncation}
\label{sectionAnElementaryExampleP2New} 

One first computes all the identities like (\ref{eqP6NormalIdentity}) 
involving
first order first degree subequations of $\PII$.
As well as for any $\Pn$, such a subequation 
can only be a Riccati equation,
which is the unique such ODE with the Painlev\'e property,
\begin{eqnarray}
& & R \equiv U' + a_2 U^2 + a_1 U + a_0=0.
\end{eqnarray}
Eliminating $U'$ with $\PII(U)$, one obtains
\begin{eqnarray}
\forall U & : & 
2(a_2^2-D^2)U^3
+(3 a_1 a_2 - a_2') U^2
\nonumber
\\
& &
+(2 a_2 a_0 + a_1^2 - a_1'-D^2 x) U
+(a_1 a_0 - a_0' - \Alpha)=0,
\label{eq32}
\end{eqnarray}
a system admitting the unique solution
($D$ denotes any square root of $D^2$)
\begin{eqnarray}
& & 
R \equiv U' + D \left(U^2 + \frac{x}{2}\right)=0,\
2 \Alpha + D =0,
\label{eqP2subeqConstraint}
\end{eqnarray}
i.e. the well known algebraic transform of an Airy equation.
The indeterminacy on $R$ (which, up to now, is only defined 
up to an additive term containing the factor $2 \Alpha + D$)
is removed by the explicit form of the 
identity between $\PII$ and its subequation,
\begin{eqnarray}
\forall (\Alpha,D) 
& & 
R'-\left(\Alpha + \frac{D}{2}\right) - (2 D U) R + \PII(U,x,\Alpha,D)\equiv 0,
\label{eqP2NormalR}
\end{eqnarray}
a relation valid for any $\Alpha$ and $D$.
By identification with (\ref{eqRiccatiDenom}) according to the proof presented
in section \ref{sectionOneFamilyTruncationHK},
one obtains
\begin{eqnarray}
& & 
g_2=D \left(U^2 + \frac{x}{2}\right)\cdot
\end{eqnarray}

Given the coefficients of the three terms of $\PII$
\begin{eqnarray}
& &
A_2=0,\
A_1=0,\
A_0= D^2 (2 U^3 + x U) + \Alpha,\
\end{eqnarray}
equation (\ref{eqg0Apriori}) then provides the value of $g_0$,
which only depends on $(A_2,A_1,A_0,g_2)$,
\begin{eqnarray}
& &
g_0= \Alpha + \frac{D}{2}\cdot
\end{eqnarray}
Therefore, before undertaking the truncation properly said,
there only remains to find $g_1$ and the two relations between
$(d,\alpha,D,\Alpha)$.

The assumption for the truncation is
\begin{eqnarray}
& & 
u=U+u_0 Z^{-1},\
u_0=d^{-1},\
\\
& &
u''=d^2 (2 u^3 + x u) + \alpha,\
U''=D^2 (2 U^3 + x U) + \Alpha,\
\\
& &
U'=-g_2 + \frac{g_0}{Z^{-1} - g_1}\ccomma\
g_2=D \left(U^2 + \frac{x}{2}\right),\
g_0= \Alpha + \frac{D}{2}\ccomma
\\
& &
Z'=1 + z_1 Z + z_2 Z^2,\ 
\\
& &
z_1=-2 g_1 + 2 D U,\
z_2= -g_1 z_1 - g_1^2 - \partial_x g_1 - (\partial_U g_1) U'.
\end{eqnarray}
The elimination of $u,Z',U'',U',g_2,g_0,z_1,z_2$ generates the
truncated Laurent series 
\begin{eqnarray}
& &
E(u) = \sum_{j=0}^{-q+2} E_j(U,x,g_1,d,\alpha,D,\Alpha) Z^{j+q-2}=0,
\label{eqST1LaurentEP2}
\end{eqnarray}
independent of $U'$,
and one requires its identical vanishing in $Z$,
\begin{eqnarray}
& &
\forall j\ :\            E_j(U,x,g_1,d,\alpha,D,\Alpha)=0.
\label{eqST1DeterminingP2}
\end{eqnarray}
Since these six determining equations must be solved
by ascending values of $j$,
let us write each of them after insertion of the solution of the
previous ones.
Introducing the notation
\begin{eqnarray}
& &
\alpha=-d \theta_\infty,\
\Alpha=-D \Theta_\infty.
\end{eqnarray}
these are
\begin{eqnarray}
E_0 & \equiv & 0,
\\
E_1 & \equiv & (d-D) U + g_1 =0, 
\\
E_2 & \equiv & (d^2-D^2)x=0,
\\
E_3 & \equiv & 1-\Theta_\infty-\theta_\infty=0,
\\
E_j & \equiv & 0,\ j=4,5.
\end{eqnarray}
One notices that the equation $E_4=0$, corresponding to the Fuchs index,
is identically satisfied,
and that there is no need to consider $j \ge 4$
(just like for $\PVI$, see Ref.~\cite{CM2001c}).
These determining equations are solved as
\begin{eqnarray}
& & 
g_1=(D-d) U,\
D^2=d^2,\
\theta_\infty=1-\Theta_\infty,
\end{eqnarray}
and the first half of the birational transformation is
\begin{eqnarray}
& & 
\frac{D(1/2-\Theta_\infty)}{d u - D U}=U'+D \left(U^2 + \frac{x}{2}\right).
\end{eqnarray}
Since, whatever be the choice of sign $D=\pm d$,
the relation between the parameters $(d,\theta_\infty,D,\Theta_\infty)$
is an involution,
the second half is obtained by just permuting the uppercase and lowercase
notation,
\begin{eqnarray}
& & 
\frac{d(1/2-\theta_\infty)}{D U - d u}=u'+d \left(u^2 + \frac{x}{2}\right).
\end{eqnarray}

Although the two choices $d=\pm D$ are equally acceptable
(the unique homography which conserves $x$, namely $(u,x) \mapsto (-u,x)$,
allows one to freely reverse the sign of $D$),
the choice $d=D$ is better because this is the one which is inherited
from $\PVI$ by the confluence \cite{CM2001c},
so the final result is the involution
\begin{eqnarray}
& & 
\frac{(\theta_\infty -\Theta_\infty)/2}{u - U}
=U'+D \left(U^2 + \frac{x}{2}\right)
=u'+d \left(u^2 + \frac{x}{2}\right),
\label{eqP2bira0}
\\
& &
d=D,\
\theta_\infty=1-\Theta_\infty.
\label{eqP2bira}
\end{eqnarray}

\textit{Remark}.
When compared to the Riccati equation (\ref{eqRiccatiOneFamily}) for $Z$,
the identity (\ref{eqP2NormalR})
shows that the values
\begin{eqnarray}
& & 
Z=\frac{U'+D(U^2+x/2)}{\Alpha+D/2},\
z_1=2 D U,\
z_2=0,
\end{eqnarray}
i.e.
\begin{eqnarray}
& & 
g_0=\Alpha+\frac{D}{2},\
g_1=0,\
g_2=D \left(U^2 + \frac{x}{2}\right),
\end{eqnarray}
define \textit{a priori} a particular solution of the truncation.
The computation has shown that, 
\textit{modulo} the homography $(u,x,\alpha) \to (-u,x,-\alpha)$,
this solution is unique.

\subsection{Comparison with the semi-truncation} 
\label{sectionAnElementaryExampleP2Old} 

As already explained in summer school lecture notes \cite{CetraroConte},
the method proposed in Ref.~\cite{GJP1999a} is in fact not distinct
from a truncation.
Therefore we will adopt the truncation language to clarify its presentation
and perform the comparison.

The assumption for the semi-truncation is 
\begin{eqnarray}
& & 
u=U+u_0 Z^{-1},\
u_0=d^{-1},\
\label{eqDTST}
\\
& &
u''=d^2 (2 u^3 + x u) + \alpha,\
U''=D^2 (2 U^3 + x U) + \Alpha,\
\\
& &
Z'=1 + z_1 Z + z_2 Z^2.
\end{eqnarray}
The elimination of $u,Z',U''$ generates the truncated Laurent series
\begin{eqnarray}
& &
E(u) = \sum_{j=0}^{-q} E_j(U',U,z_1,z_2,\veca,\vecA,x) Z^{j+q}=0,\
\veca=(\alpha,d),\
\vecA=(\Alpha,D),
\label{eqST1LaurentEwithUprime}
\end{eqnarray}
in which the dependence on $U'$ has been emphasized,
and one does \textit{not} require its identical vanishing in $Z$.
Indeed, the four coefficients of this Laurent series in $Z$ are
\begin{eqnarray}
E_0 & \equiv & 0,\
\\
E_1 & \equiv & 3 (2 d U - z_1),
\\
E_2 & \equiv & z_1' - z_1^2 - 2 z_2 + d^2 (6 U^2 + x),
\\
E_3 & \equiv & z_2' - z_1 z_2 + d (d^2-D^2) (2 U^3 + x U)+d (\alpha - \Alpha).
\end{eqnarray}

The first two coefficients $E_j,j=0,1$,
even for cases other than $\PII$,
are independent of $U'$ 
and the same in the two truncations
if one remembers the correspondence (\ref{eqz1Apriori})
between $z_1$ and $(g_j,A_j)$.
Therefore, defining the two equations $E_j=0,j=0,1$ and solving them
makes no difference betwen the two methods and provides
\begin{eqnarray} 
& &
z_1=2 d U.
\end{eqnarray}

As soon as $j \ge 2$,
the coefficients $E_j$ are essentially different in the two truncations,
and this is because the remaining truncated Laurent series
depends on both $U'$ and $Z$,
\begin{eqnarray}
& &
\sum_{j=2}^{-q}  E_j (U',U,z_2,\veca,\vecA,x) Z^{j+q}=0,
\label{eqST1LaurentERemainder}
\end{eqnarray}
while they are not independent but linked by the unused relation
(to be found \textit{in fine})
(\ref{eqHomographyRUprime}).

Solving the system $E_j=0,j=0,\ldots,3$, would indeed yield \cite{GNTZ}
\begin{eqnarray} 
& &
z_1=2 d U,\
z_2=d U' + (d U)^2 + \frac{1}{2} d^2 x,\
\alpha + \frac{d}{2}=0,
\end{eqnarray}
which cannot define a \BiT.

To overcome this first difficulty,
after solving $E_1=0,E_2=0$ (hence the name semi-truncation),
one eliminates $Z$ between 
(\ref{eqRiccatiOneFamily}) and
(\ref{eqST1LaurentERemainder}),
which amounts to compute the resultant of two polynomials of $Z$
and generically results in
\begin{eqnarray}
& &
F(U',U;z_2,\veca,\vecA,x) =0,
\label{eqST1F}
\\
& &
Z=z(U',U;z_2,\veca,\vecA,x),
\label{eqST1Z}
\end{eqnarray}
in which $F$ is a differential polynomial 
and $z$ a rational function of their arguments.
The two equations (\ref{eqDTST}) and (\ref{eqST1Z}) will define the 
first half of a \BiT,
not necessarily of degree one,
after the first equation (\ref{eqST1F}) has been solved for $z_2$.

Solving (\ref{eqST1F}) for $z_2$ is the second difficulty of the method.
Indeed, the method is to enforce the irreducibility of $\Pn$
by requiring the identical vanishing of (\ref{eqST1F}) as a
polynomial of $U',U$.
But (\ref{eqST1F}) depends on $z_2,z_2',z_2''$
and this procedure first requires an additional assumption on the
explicit dependence of $z_2$ on $(U',U)$.
The assumption \cite{GJP1999a}
\begin{eqnarray}
& &
z_2=f_0(x) + f_1(x) U
\end{eqnarray}
is sufficient for $\PII$ and it leads to the expected result
(\ref{eqP2bira0})--(\ref{eqP2bira}).

Finally, there exists a third difficulty,
which only occurs for $\PVI$,
this is the value $1$ of the Fuchs index of $\PVI$,
a value which cannot be changed by homography on $u$.
In this case, the coefficient $E_1$ is identically zero
and does not determine $z_1$,
so one must make two assumptions for the dependence of $(z_1,z_2)$
on $(U',U)$.
This is why, to our knowledge,
the semi-truncation method has not been applied to $\PVI$ yet.

\section{Processing of $\PV$ with the present truncation} 
\label{sectionImprovedTruncationP5}

$\PVI$ has already been processed with our method 
\cite{CM2001alreadysubmitted,CM2001c} 
before the present improvement,
and the solution to the truncation is unique. 
The first reason for choosing $\PV$ here is that,
as opposed to $\PVI$,
one expects at least two inequivalent solutions,
a situation which only occurs for $\PV$ and $\PIV$.
The second reason is to show how the difficulty arising from the index $1$
is overcome.
These two inequivalent first degree {\BiT}s for $\PV$ were first found
by Gromak \cite[Eq.~(13)]{Gromak1976}
and by Okamoto \cite{Okamoto1987II}.

The definition of $\PV$ is
\begin{eqnarray*}
\PV\ : \
u''
&=&
\left[\frac{1}{2 u} + \frac{1}{u-1} \right] {u'}^2
- \frac{u'}{x}
+ \frac{(u-1)^2}{x^2} \left[ \alpha u + \frac{\beta}{u} \right]
+ \gamma \frac{u}{x}
+ \delta \frac{u(u+1)}{u-1}\ccomma
\nonumber
\end{eqnarray*}
and 
the data for the two opposite families of movable singularities of $u$ are
\begin{eqnarray}
& &
p=-1,\
q=-5,\
u_0^2=\theta_\infty^{-2} x^2,\ 
\alpha=\theta_\infty^2/2,\
\hbox{ Fuchs index } 1,
\end{eqnarray}
with the definition of the monodromy exponents 
\cite{Okamoto1986Pn}, 
\begin{eqnarray}
& &
\theta_\infty^2= 2 \alpha,\
\theta_0^2= - 2 \beta,\
d \theta_1= - \gamma,
d^2=-2 \delta.
\end{eqnarray}

The search for the Riccati subequations
as explained in Section \ref{sectionAnElementaryExampleP2New}
($f_\infty,f_0,f_1$ denote the three functions of $x$ to be found)
\begin{eqnarray}
& &
R=\frac{x U'}{U (U-1)^2}
 +\frac{f_0}{U}+\frac{f_\infty-f_0}{U-1}+\frac{f_1 x}{(U-1)^2}\ccomma
\end{eqnarray}
leads to the unique algebraic solution
\begin{eqnarray}
& &
f_0^2=\Theta_0^2,\
f_\infty^2=\Theta_\infty^2,\
f_1^2=D^2,\
(1+f_\infty-f_0) f_1 = D \Theta_1.
\end{eqnarray}
After choosing the square roots,
there are two distinct identities of the type
\begin{eqnarray}
& &
R' + F_2(U,x) R^2 + F_1(U,x) R + F_0(U,x) + \PV(U) \equiv 0,
\end{eqnarray}
namely
\begin{eqnarray}
{\hskip -1.2truecm}
& &
R=\frac{x U'}{U (U-1)^2}
 +\frac{\Theta_0}{U}+\frac{\Theta_1-1}{U-1}+\frac{D x}{(U-1)^2}\ccomma
\label{eqP5NormalR}
\\
{\hskip -1.2truecm}
& &
x R'
+\frac{1}{2}(1-\Theta_\infty-\Theta_0-\Theta_1)
            (1+\Theta_\infty-\Theta_0-\Theta_1)
+\frac{1}{2}\left((U-1)^2+2 U(U-1)\right) R^2
\nonumber \\ 
{\hskip -1.2truecm}
& & \phantom{x R'}
-\left((\Theta_1-1)U+(2\Theta_0+\Theta_1-1)(U-1)+d x\right) R
+\frac{x^2}{U^2(U-1)^2}\PV(U)\equiv 0,
\label{eqP5NormalIdentity}
\end{eqnarray}
and
\begin{eqnarray}
{\hskip -1.2truecm}
& &
R=\frac{x U'}{U (U-1)^2}
 +\frac{\Theta_0}{U}+\frac{\Theta_\infty-\Theta_0}{U-1}+\frac{D x}{(U-1)^2}
\ccomma
\label{eqP5BiasedR}
\\
{\hskip -1.2truecm}
& &
x R'
- \frac{D (1+\Theta_\infty-\Theta_0-\Theta_1)x}{(U-1)^2}
+\frac{1}{2}\left((U-1)^2+2 U(U-1)\right) R^2
\nonumber \\ 
{\hskip -1.2truecm}
& & \phantom{x R'}
-\left((\Theta_\infty-\Theta_0)U+(\Theta_\infty+\Theta_0)(U-1)+d x\right) R
+\frac{x^2}{U^2(U-1)^2}\PV(U)\equiv 0.
\end{eqnarray}
The reason why they are essentially distinct is the two different 
factorizations of the condition on $(\Theta_\infty,\Theta_0,\Theta_1,D)$.
As shown in Ref.~\cite{CM2001c},
these two cases are inherited from the first degree \BiT\ of Okamoto for 
$\PVI$ by two different confluences,
which have been called respectively \textit{normal}
or \textit{unbiased}, and \textit{biased} in Ref.~\cite{CM2001c}.

Each choice will correspond to a different solution to the truncation,
defining two distinct first degree {\BiT}s,
respectively denoted $\TPVu$ and $\TPVb$ (like unbiased, biased).
We now follow exactly the steps of Section 
\ref{sectionAnElementaryExampleP2New}.

\subsection{The normal (or unbiased) \BiT\ of $\PV$}

The first possibility (\ref{eqP5NormalR}),
\begin{eqnarray}
& &
g_2=
\frac{U(U-1)^2}{x}
\left(\frac{\Theta_0}{U}+\frac{\Theta_1-1}{U-1}+\frac{D x}{(U-1)^2}\right)
\ccomma
\end{eqnarray}
provides the factorization 
(compare with the identity (\ref{eqP5NormalIdentity}))
\begin{eqnarray}
& &
g_0=
-(1-\Theta_\infty-\Theta_0-\Theta_1)(1+\Theta_\infty-\Theta_0-\Theta_1)
\frac{U(U-1)^2}{2 x^2}\cdot
\end{eqnarray}
Denoting the residue $u_0$ as
\begin{eqnarray}
& &
u_0=-\theta_\infty^{-1} x,
\end{eqnarray}
the first determining equations are
\begin{eqnarray}
E_j & \equiv & 0,\ j=0,1,
\\
E_2 & \equiv & x^2 (\partial_x g_1 -g_1^2) -x g_2 \partial_U g_1
\nonumber
\\
& &
+x \left(
2 (\theta_\infty -1 + \Theta_0 + \Theta_1) U
+ 2 -2 \Theta_0 - \Theta_1 -\frac{4}{3}\theta_\infty + D x
\right) g_1
\nonumber
\\
& &
+K_0 U^2 + K_1 U + \frac{d^2-D^2}{6} x^2 + K_2 x + K_3=0,
\end{eqnarray}
in which the $K_m$'s are constants.
The reason for the identical vanishing of $E_1$ is the Fuchs index $1$,
but this does not harm the computation.
Indeed, the only possibility for $g_1$ to be rational in $U$ is that it be
a first degree polynomial of $U$,
which the singularities of $\PV$ in $U$ suggest to define as
\begin{eqnarray}
& & g_1=\frac{f_0(x) U +(f_1(x)-f_0(x)) (U-1)}{x}\cdot
\label{eqg1def}
\end{eqnarray}
Equation $E_2=0$ then splits into
\begin{eqnarray}
& & E_2 \equiv \sum_{k=0}^{2} E_2^{(k)} U^k,\
\forall k\ E_2^{(k)}(f_0,f_1,x,\vect,\vecT)=0,
\label{eqsystem3}
\\
& & E_2^{(2)}\equiv
\Theta_\infty^2-\left(2 f_1-2\theta_\infty +1-\Theta_0-\Theta_1\right)^2=0.
\end{eqnarray}
This system (\ref{eqsystem3}) of three equations is equivalent to
\begin{eqnarray}
& &
g_1=(2\theta_\infty-1-\Theta_\infty+\Theta_0+\Theta_1)\frac{3 U-2}{6 x}
\ccomma\
d^2=D^2,
\\
& & 6 \theta_0^2 -2 \theta_\infty^2=
\left(2 - \Theta_\infty+\Theta_0-2 \Theta_1\right)^2 - 3 (\Theta_1-1)^2,
\end{eqnarray}
and there only remains to find one algebraic relation between
the monodromy exponents.

Since all the dependence on $U$ is now found,
the next determining equation splits even more,
according to both powers of $U$ and $x$,
\begin{eqnarray}
{\hskip -1.0truecm}
& & E_3 \equiv (2\theta_\infty-1-\Theta_\infty+\Theta_0+\Theta_1)
\left[
E_3^{(1,0)} U + \left(D^2 x^2 + E_3^{(0,1)} x + E_3^{(0,0)}\right)
\right]
=0,\
\end{eqnarray}
a system which admits as only solution the vanishing of the first factor.
Therefore, the coefficient $g_1$ vanishes,
like in the whole normal sequence \cite{CM2001c}.
This ends the resolution, therefore achieved at $j=3$,
just like in the trunction for $\PVI$ \cite{CM2001alreadysubmitted}.

This first solution $\TPVu$ has the affine representation
\begin{eqnarray}
& &
\PV\ : \
\pmatrix{\theta_\infty \cr
         \theta_0      \cr
         \theta_1      \cr}
= \frac{1}{2} \pmatrix {1 & -1 & -1 \cr -1 & 1 & -1 \cr -2 & -2 & 0 \cr}
\pmatrix{\Theta_\infty \cr
         \Theta_0      \cr
         \Theta_1      \cr}
              + \frac{1}{2} \pmatrix{ 1 \cr 1 \cr 2 \cr}\ccomma\
d = D,
\end{eqnarray}
in which the six arbitrary signs of $\vect$ and $\vecT$ are chosen 
in such a way that the square of this transformation is the identity.
The birational representation is
\begin{eqnarray}
\PV\ : \
\frac{1-\Theta_\infty-\Theta_0-\Theta_1}{u-U}
& = &
  \frac{x U'}{U (U-1)^2}
 +\frac{\Theta_0}{U}+\frac{\Theta_1-1}{U-1}+\frac{D x}{(U-1)^2}\cdot
\end{eqnarray}

This transformation for $\PV$ was first found by Okamoto \cite{Okamoto1987II}.

For completion, the values of $z_1,z_2$ are
\begin{eqnarray}
& & z_1= \frac{\Theta_1 U +(2 \Theta_0 + \Theta_1-2) (U-1)}{x}\ccomma
\\
& & z_2= 
(-\Theta_\infty+\Theta_0+\Theta_1-1)(\Theta_\infty+\Theta_0+\Theta_1-1)
\frac{2 U (U-1) + (U-1)^2}{4 x^2}\cdot
\end{eqnarray}

\subsection{The biased \BiT\ of $\PV$}

With the second possibility (\ref{eqP5BiasedR}),
\begin{eqnarray}
& &
g_2=
\frac{U(U-1)^2}{x}
\left(
\frac{\Theta_0}{U}+\frac{\Theta_\infty-\Theta_0}{U-1}+\frac{D x}{(U-1)^2}
\right)\ccomma
\end{eqnarray}
one similarly obtains
\begin{eqnarray}
& &
g_0=(1+\Theta_\infty-\Theta_0-\Theta_1)\frac{D U}{x}\cdot
\end{eqnarray}
In order to later make easier our involution convention,
it is convenient this time to denote the residue $u_0$ as
\begin{eqnarray}
& &
u_0=\theta_\infty^{-1} x.
\end{eqnarray}
The first three determining equations are
\begin{eqnarray}
{\hskip -8.0 truemm}
E_j & \equiv & 0,\ j=0,1,
\\
{\hskip -8.0 truemm}
E_2 & \equiv & -x^2 (U-1) (\partial_x g_1 -g_1^2) + x (U-1) g_2 \partial_U g_1
\nonumber
\\
{\hskip -8.0 truemm}
& &
+x \left(
\frac{2}{3}\theta_\infty (U-1)(3 U+2)+(\Theta_\infty-\Theta_0) (U-1)+D x (U+1)
\right) g_1
\nonumber
\\
{\hskip -8.0 truemm}
& &
+\theta_\infty(\theta_\infty-\Theta_\infty) U^3
\nonumber
\\
{\hskip -8.0 truemm}
& &
+K_0 U^2 - \frac{d^2-D^2}{6} x^2 (U-1) + (K_1 x + K_2) U + K_3 x + K_4=0.
\end{eqnarray}
Again,
the only possibility for $g_1$ to be rational is
to be a first degree polynomial of $U$, defined as in (\ref{eqg1def}),
and equation $E_2=0$ is equivalent to
\begin{eqnarray}
& &
g_1=- \theta_\infty \frac{3 U-2}{2 x}
 -\frac{\Theta_\infty-\Theta_0-\Theta_1+1}{6 x}
\ccomma\
d^2=D^2,
\\
& & 6 \theta_0^2 -2 \theta_\infty^2=
\left(- \Theta_\infty+\Theta_0-2 \Theta_1-1\right)^2 - 3 \Theta_1^2.
\end{eqnarray}
Last, equation $E_3=0$ yields
\begin{eqnarray}
\theta_\infty=\frac{-\Theta_\infty+\Theta_0+\Theta_1-1}{2}\cdot
\end{eqnarray}

This second solution $\TPVb$ is represented by the involution
\begin{eqnarray}
{\hskip -8.0 truemm}
& &
\PV\ : \
\pmatrix{\theta_\infty \cr
         \theta_0      \cr
         \theta_1      \cr}
=-\frac{1}{2} \pmatrix {1 & -1 & -1 \cr -1 & 1 & -1 \cr -2 & -2 & 0 \cr}
\pmatrix{\Theta_\infty \cr
         \Theta_0      \cr
         \Theta_1      \cr}
+ \frac{1}{2} \pmatrix { -1 \cr 1 \cr 0 \cr},\
d = - D,
\end{eqnarray}
and
\begin{eqnarray}
{\hskip -8.0 truemm}
\PV
& : &
\frac{-2 D x}{(u-1)(U-1)}
 =
(U-1) \left(
  \frac{x U'}{U (U-1)^2}
 +\frac{\Theta_0}{U}+\frac{\Theta_\infty-\Theta_0}{U-1}+\frac{D x}{(U-1)^2} 
\right)
\ccomma\
\end{eqnarray}
with
\begin{eqnarray}
z_1 
& = &
\frac{(\Theta_\infty-\Theta_0)U+(2\Theta_0+\Theta_1-1)(U-1)}{x}
+ \frac{2 U'}{U-1}\ccomma
\label{eqP5z1withUprime}
\\
\frac{2 x^2}{\theta_\infty} z_2
& = &
6 x U' + 2 D x \left(\frac{2}{U-1}+U+2\right)
\nonumber
\\
& &
+2 (\Theta_\infty-\Theta_0-1) U(U-1)
+  (\Theta_\infty+3\Theta_0+\Theta_1+1) (U-1)^2.
\label{eqP5z2withUprime}
\end{eqnarray}

This second transformation has first been obtained 
by Gromak \cite[Eq.~(13)]{Gromak1976}.

Let us denote $\H$ the unique homography of $\PV$ which conserves $x$,
\begin{eqnarray}
\PV
& : &
\H (x,u     ,\theta_\infty,\theta_0,\theta_1)
=  (x,u^{-1},\theta_0,\theta_\infty,\theta_1),
\end{eqnarray}
and $\Sa,\Sb,\Sc$ the operators which reverse the sign of, respectively,
$\theta_\infty,\theta_0,\theta_1$.
One has the relation
\begin{eqnarray}
& &
\TPVu=\Sa \TPVb \Sa \Sc \TPVb \Sa \H,
\end{eqnarray}
but we could not find an inverse relation expressing the 
biased transformation as powers of the unbiased one.
Therefore, $\TPVb$ is more elementary than $\TPVu$.

Let us compare again with the semi-truncation.
In \cite{GJP2001N},
to avoid the third difficulty mentioned in Section 
\ref{sectionAnElementaryExampleP2Old},
the authors first change the Fuchs index to $2$ by performing
a homography on $\PV$.
The second difficulty is handled with the extra assumption that
$z_2$ (their $\tau$) should be independent of $U'$.
This then allows them to obtain the unbiased transformation $\TPVu$.
The reason why they fail to find the biased one $\TPVb$ with the
semi-truncation
is the restricting assumption on $z_2$,
since in this case $z_2$ explicitly depends on $U'$,
see expression (\ref{eqP5z2withUprime}).
By looking at the ODE satisfied by $Z$
(evidently an algebraic transform of $\PV$),
they finally obtain this second missing first degree \BiT.

\section{Conclusion} 

The improvement which we have presented to the truncation
drastically reduces the amount of computation,
and the search for first degree {\BiT}s of higher order ODEs 
by this method becomes much easier.
This will be addressed in future work.

\section*{Acknowledgments} 

The authors are grateful to the organizers for their financial support
to attend the conference.
They acknowledge the financial support of the Tournesol grant T99/040.
MM acknowledges the financial support of
the IUAP Contract No.~P4/08 funded by the Belgian government
and the support of CEA.
This work was also performed in the framework of the INTAS project 99-1782.


\label{lastpage}


\begin{thebibliography}{99}
\small

\bibitem{CJP} P.~A.~Clarkson, N.~Joshi, and A.~Pickering,
B\"acklund transformations for the second Painlev\'e hierarchy:
a modified truncation approach,
Inverse Problems {\bf 15} (1999) 175--187.

\bibitem{GJP1999a} P.~Gordoa, N.~Joshi, and A.~Pickering,
Mappings preserving locations of movable poles:
a new extension of the truncation method to ordinary differential equations,
Nonlinearity {\bf 12} (1999) 955--968.

\bibitem{CM2001alreadysubmitted} R.~Conte and M.~Musette,
New contiguity relation of the sixth Painlev\'e equation from a truncation,
13 pages, preprint S2001/009, submitted (2001). 

\bibitem{CM2001c} R.~Conte and M.~Musette, 
First degree birational transformations of the Painlev\'e equations
and their contiguity relations,
J.~Phys.~A {\bf ?} (2001) ?--?. nlin.SI/0110028.
Special issue SIDE IV Proceedings, Tokyo, 27 November--1 December 2000. 

\bibitem{C2001TSP6} R.~Conte,
Sur les transformations de Schlesinger de la sixi\`eme \'equation de 
Painlev\'e,
\CRAS\ {\bf 332} (2001) 501--504. math.CA/0103165.

\bibitem{Okamoto1987I} K.~Okamoto,
Studies on the Painlev\'{e} equations, 
I, Sixth Painlev\'e equation, Ann.~Mat.~Pura Appl.~{\bf 146} (1987) 337--381.

\bibitem{WTC} J.~Weiss, M.~Tabor, and G.~Carnevale,
The Painlev\'e property for partial differential equations,
J.~Math.~Phys.~{\bf 24} (1983) 522--526.
 
\bibitem{Cargese1996Musette} M.~Musette,
Painlev\'e analysis for nonlinear partial differential equations,
{\it The Painlev\'e property, one century later}, 
517--572,
ed.~R.~Conte,
CRM series in mathematical physics (Springer, New York, 1999).

\bibitem{CetraroConte} R.~Conte,
Exact solutions of nonlinear partial differential equations
by singularity analysis,
\textit{Direct and inverse methods in nonlinear evolution equations},
83 pages,
ed.~A.~Greco (Springer, Berlin, 2001).
nlin.SI/0009024.
CIME school, Cetraro, 5--12 September 1999.

\bibitem{PaiLecons} P.~Painlev\'e,
{\it Le\c{c}ons sur la th\'eorie analytique des \'equations diff\'erentielles}
(Le\c{c}ons de Stockholm, 1895)
(Hermann, Paris, 1897).
Reprinted, {\it \Oeuvres\ de Paul Painlev\'e}, vol.~I
(\'Editions du CNRS, Paris, 1973).
 
\bibitem{FuchsP6} R.~Fuchs,
Sur quelques \'equations diff\'erentielles lin\'eaires du second ordre,
\CRAS\ {\bf 141} (1905) 555--558. 

\bibitem{GNTZ} J.~D.~Gibbon, A.~C.~Newell, M.~Tabor, and Zeng Y.-b,
Lax pairs, B\"acklund transformations and special solutions for ordinary
differential equations,
Nonlinearity {\bf 1} (1988) 481--490.

\bibitem{Gromak1976} V.~I.~Gromak,                              
Solutions of Painlev\'e's fifth problem,
Differentsial'nye Uravneniya {\bf 12} (1976) 740--746
       [English~: Diff.~equ.~{\bf 12} (1976) 519--521].

\bibitem{Okamoto1987II} K.~Okamoto,                                
Studies on the Painlev\'{e} equations, 
II, Fifth Painlev\'e equation, Japan.~J.~Math., {\bf 13} (1987) 47--76.

\bibitem{Okamoto1986Pn} K.~Okamoto,                                    
Isomonodromic deformation and Painlev\'e equations, and the Garnier system,
J.~Fac.~Sci.~Univ.~Tokyo, Sect.~IA {\bf 33} (1986) 575--618.

\bibitem{GJP2001N} P.~Gordoa, N.~Joshi, and A.~Pickering,          
Mappings preserving locations of movable poles:
II. The third and fifth Painlev\'e equations,
Nonlinearity {\bf 14} (2001) 567--582.

\end{thebibliography}
\end{document}